# Ferromagnetic spin fluctuation in LaFeAsO$_{1-x}$F$_x$

Yoshimitsu Kohama[1], Yoichi Kamihara[2], Masahiro Hirano[2,3], Hitoshi Kawaji[1], Tooru Atake[1], and

Hideo Hosono[1–3]

[1]*Materials and Structures Laboratory, Tokyo Institute of Technology 4259 Nagatsuta-cho, Midori-ku, Yokohama 226-8503, Japan*
[2]*ERATO-SORST, Japan Science and Technology Agency in Frontier Research Center, Tokyo Institute of Technology, 4259 Nagatsuta-cho, Midori-ku, Yokohama 226-8503, Japan*
[3]*Frontier Research Center, Tokyo Institute of Technology,* 4259 Nagatsuta-cho, Midori-ku, Yokohama 226-8503, Japan

The F-doped LaFeAsO, a recently discovered superconductor with the high $T_c$ of 26 K, has been studied by the resistivity ($\rho$), magnetic susceptibility ($\chi$), and heat capacity ($C_p$) measurements in the F doping range $0 \leq x \leq 0.14$ (LaFeAsO$_{1-x}$F$_x$). In the low temperature region, a $T^3\ln T$ term in $C_p$ and a $T^2$ term in $\chi$, which are derived from the spin fluctuation, are observed. The nearly ferromagnetic nature evidenced by a large Wilson ratio (6.5 for $x = 0$, and 11.2 for $x = 0.025$) suggests that the superconductivity in the LaFeAsO system is mediated by ferromagnetic spin fluctuation.

Since the discovery of high-temperature superconductivity in cuprate oxide compounds [1], a number of experimental and theoretical studies have been devoted to understanding the superconducting mechanism [2-4]. The most widely accepted physical picture for explaining such a high transition temperature ($T_c$) is that the antiferromagnetic (AF) spin fluctuation gives rise to the strong pair interaction of the Cooper pairs [2,3]. So far, it has been demonstrated that this mechanism is successfully realized in various superconductors such as the Pu-based compound PuCoGa$_5$ ($T_c$ = 18.5 K) [5], organic compounds $\kappa$-(BEDT-TTF)$_2$X ($T_c \sim 10$ K) [6], and so on. In contrast, there are only a few superconductors mediated by ferromagnetic (FM) spin fluctuation, and these show significantly low $T_c$ [7,8]. Therefore, a system having FM spin fluctuation was believed to be an unlikely candidate for producing high-$T_c$.

Recently, we found that aliovalent ion doping of fluorine into a layered oxyarsenide LaFeAsO produces superconductivity with a high transition temperature of $T_c \sim 26$ K [9] and a high upper critical field [10]. LaFeAsO has a layered structure belonging to the ZrCuSiAs structure with space group $P4/nmm$ [10]. Co [11], Fe [9,12], and Ni-based [13] analog compounds show metallic conductivity without carrier doping. Fe and Ni-based compounds exhibit superconductivity, and Co-based compounds have FM metal properties (e.g., LaCoPO [11] and LaCoAsO [11] having Curie temperatures of 43 and 66 K, respectively). The occurrence of itinerant ferromagnetism means that the interaction between quasiparticles can be considered to be FM, and the presence of FM spin fluctuation may be expected in the LaFeAsO system.

If the presence of FM spin fluctuation in this system is demonstrated, it would be evidence that the high-$T_c$ in the system is mediated by FM spin fluctuation. In general, FM spin fluctuation effects manifest themselves in a $T^2$ term in susceptibility [14] and $T^3\ln(T/T_{SF})$ term in heat capacity [15], well below the characteristic temperature of spin fluctuation ($T_{SF}$). In this letter, we present the results of resistivity, magnetic susceptibility, and heat capacity measurements in LaFeAsO$_{1-x}$F$_x$ ($x$ =

0, 0.025, 0.05, 0.11, and 0.14), and show the characteristic properties of FM spin fluctuation.

Polycrystalline samples of LaFeAsO$_{1-x}$F$_x$ ($x$ = 0, 0.025, 0.05, 0.11 and 0.14) were prepared by solid-state reactions of LaAs, FeAs, Fe$_2$As, La, LaF$_3$, and La$_2$O$_3$ powders in an evacuated quartz tube, as reported previously [9]. The value of $x$ was estimated from the Vegard's law [9]. Powder X-ray diffraction showed that the samples with $x$ = 0.025, 0.11 and 0.14 contained impurities of FeAs (~3%) and LaOF (~3%). Samples with $x$ = 0 and 0.05 contained only a small amount of FeAs (~1%).

Resistivity, magnetic susceptibility, and heat capacity measurements were performed between 2 and 300 K using a Physical Property Measurement System (PPMS) from Quantum Design Inc.

Figure 1 shows the heat capacity ($C_p$) and magnetic susceptibility ($\chi$) of LaFeAsO$_{1-x}$F$_x$ ($x$ = 0, 0.025, 0.05, 0.11, and 0.14). A Sharp drop in $\chi$ corresponding to superconducting transition is observed for $x$ = 0.05, 0.11, and 0.14. For $x$ = 0 and 0.025, no such drop appears until 2 K, and another anomaly is observed at ~150 K, as also seen in $C_p$ and resistivity ($\rho$) [10]. The onset temperatures of the diamagnetic signal are consistent with the peak temperatures of the differential resistance [10]. The bulk nature of superconductivity for $x$ = 0.05 and 0.11 is now confirmed by the present $C_p$ measurement. The inset in Fig. 1(a) shows the heat capacity difference ($C_p^*$) between superconducting samples ($x$ = 0.05, 0.11, and 0.14) and non-superconducting LaFeAsO below 35 K; i.e., the difference in electronic contribution, $C_p^* = (C_{ele}(x) + C_{lat}(x)) - (C_{ele}(x=0) + C_{lat}(x=0)) \sim C_{ele}(x) - C_{ele}(x=0)$, where $C_{ele}$ and $C_{lat}$ are the electronic and the lattice contribution, respectively. Taking into account entropy conservation between the superconducting and normal states, the heat capacity jump at $T_c$ ($\Delta C_p^*/T_c$) is estimated to be 6.2 mJ mol$^{-1}$ K$^{-2}$ ($T_c$ = 20.5 K) and 6.4 mJ mol$^{-1}$ K$^{-2}$ ($T_c$ = 21.1 K) for $x$ = 0.05 and 0.11, respectively. For $x$ = 0.14, the heat capacity jump is small and smears out, implying the small superconducting volume fraction. In fact, the diamagnetism signal of this sample is one-third to one-fifth of that of $x = 0.05$ or $0.11$, as shown in the inset of Fig. 1(b).

Figure 2(a) shows $C_p$ values below 8 K in the plot of $C_p/T$ versus $T^2$. For $x = 0.05$, $0.11$, and $0.14$, the data points lie on straight lines with finite intercepts at $T = 0$, indicating the following temperature dependence, $C_p = \gamma_{SC} T + \beta T^3$, where $\gamma_{SC}$ is the electronic heat capacity coefficient suppressed by the occurrence of a superconducting gap and $\beta$ is the lattice heat capacity coefficient. The finite values of $\gamma_{SC}$ for the superconducting samples would be attributed to the presence of impurities, as observed in various systems [16,17]. The Debye temperature ($\Theta_D$) estimated from $\beta$ is 319 K ($x = 0.05$), 308 K ($x = 0.11$), and 332 K ($x = 0.14$), which is lower than that of LaFePO ($\Theta_D = 371$ K [16]), probably because of the heavier mass of As compared to P.

For $x = 0$ and $0.025$, $C_p/T$ distinctly departs from linear dependence below 6 K. This temperature dependence can be fitted with the characteristic $T^3 \ln(T/T_{SF})$ term of spin fluctuation system, and the total heat capacity can then be written as

$$C_p = \gamma_{SF} T + \beta T^3 + \delta T^3 \ln(T/T_{SF}), \qquad (\text{Eq.1})$$

where $\gamma_{SF}$ represents the electronic heat capacity coefficient enhanced by spin fluctuation and $\delta$ is the coefficient of the spin fluctuation term [15]. Assuming that the average of $\beta$ in these superconducting samples is equal to that in non-superconducting samples ($x = 0$ and $0.025$), Eq. 1 yields $T_{SF} = 12 \pm 1$ K and $T_{SF} = 13 \pm 1$ K for $x = 0$ and $x = 0.025$, respectively, where the error bars on $T_{SF}$ are estimated from the standard deviation of $\beta$. Figure 2(b) displays the $\chi$ versus $T^2$ plots of $x = 0$ and $0.025$ samples in the low-temperature region. $\chi$ shows a linear dependence on $T^2$ rather than the Curie-Weiss law. Béal-Monod et al. [14] reported that the characteristic temperature dependence of $\chi$ in the FM spin fluctuation system can be approximated as $\chi = \chi_0(1 - (3.2\pi^2/24)(T/T_{SF})^2)$, where $\chi_0$ is the magnetic susceptibility at 0 K. Following this expression, we obtain $T_{SF}$ of $15 \pm 1$ K and $22 \pm 1$ K for $x = 0$ and $x = 0.025$, respectively. Considering the fact that the difference in $T_{SF}$ between $C_p$ and $\chi$ measurements has been reported in previous studies [18,19], $T_{SF}$ values derived from these two independent measurements appear to be consistent with each other. In any case, the two measurements clearly demonstrate the presence of FM spin fluctuation in $x = 0$ and 0.025 samples.

Here, it is noted that the extrapolation of the high-temperature linear part in $C_p/T$ gives an electronic heat capacity coefficient without the spin fluctuation effect ($\gamma$) [20]. The values of $\gamma$ obtained from this method are $1.6 \pm 1$ and $3.2 \pm 1$ mJ K$^{-2}$ mol$^{-1}$ for $x = 0$ and 0.025, respectively. Assuming that the lattice contribution of LaFeAsO is the same as those of LaFeAsO$_{1-x}$F$_x$, the $C_p^*/T$ above the $T_c$ (see inset in Fig. 1(a)) corresponds to the difference in $\gamma$, $C_p^*/T = \gamma(x) - \gamma(x = 0)$. This assumption allows us to estimate $\gamma = 8.4 \pm 2$, $5.1 \pm 2$, and $3.8 \pm 2$ mJ K$^{-2}$ mol$^{-1}$ for $x = 0.05$, 0.11, and 0.14, respectively, where the error bar takes into account the uncertainties in $C_p^*/T$ above $T_c$ and $\gamma$ of the $x = 0$ sample. Using these values, the normalized heat capacity jump ($\Delta C_p^*/\gamma T_c$) is estimated as 0.60–0.97 and 0.90–2.0 for $x = 0.05$ and 0.11, respectively. Although the large error bars in the data do not allow a conclusive interpretation of superconductivity, the value for $x = 0.05$ is smaller than the weak-coupling BCS value of 1.43.

The electronic density of states ($N_D$) can be calculated from $\gamma$ using the relation of $\gamma = (1/3)\pi^2 k_B^2 N_D$, where $k_B$ is the Boltzmann constant. This equation gives $N_D = 0.7$, 1.4, 3.6, 2.2, and 1.6 states/eV for $x = 0$, 0.025, 0.05, 0.11, and 0.14, respectively, which are smaller than that of LaFePO (4.5 states/eV, $T_c = 3.3$ K) [16]. The smaller $N_D$ and $\Theta_D$ values in LaFeAsO should lead to lower $T_c$ within the framework of the phonon-mediated BCS theory [21], which is inconsistent with higher $T_c$. On the other hand, $\chi$ in a metal measures the density of states enhanced by the FM spin fluctuations via $\chi = \mu_B^2 N_D^*$, where $\mu_B$ is the Bohr magneton and $N_D^*$ is the enhanced density of states. The $N_D^*$ values calculated from the data at 300 K (12, 20, 31, 20, and 12 states/eV for $x = 0$, 0.025, 0.5, 0.11, and 0.14, respectively) are one order of magnitude larger than $N_D$ obtained from $C_p$. This large enhancement can be understood in terms of the FM spin fluctuation.

The ratio of $\chi$ to $\gamma$ at 0 K, which is known as the Wilson ratio ($R_W$), is commonly used as a measure to assess the electric and magnetic correlation between quasi-particles. In general, the ratio gives a dimensionless value of $R_W = 1$ for free electron systems and $R_W = 2$ for strongly correlated systems. Nearly FM systems show very large values of $R_W$ (e.g., Pd, $R_W$ = 6–8 [22]; TiBe$_2$, 12 [22]; and Sr$_3$Ru$_2$O$_7$, 10 [23]) because of the enhancement of $\chi$. In the present compounds, we obtain significantly large $R_W$ values of 6.5 and 11 for $x = 0$ and 0.025, respectively [18], indicating that LaFeAsO is considered to be a system containing FM spin fluctuation. These values are two to three times larger than that of the isostructural oxyphosphide LaFePO ($R_W$ = 4.6, $T_c$ = 3.3 K) [16], implying that the stronger spin fluctuation induces a larger Wilson ratio, and leads to enhancement of $T_c$. Such a relationship between $T_c$ and $R_w$ was found in the pressure dependence of superfluid transition in $^3$He [24].

For $x = 0$ and 0.025, $C_p$ and $\chi$ show an anomaly at about 150 K, as shown in Fig. 1. To estimate the excess heat capacity ($C_{ex}$) associated with the anomaly, we assume a smoothly varying background [10] and subtract it from the total measured $C_p$. Figure 3(a) shows the $C_{ex}$ value near the anomaly. Apparently, two peaks are present at 142 and 153 K for $x = 0$. In contrast, there is only one peak at 134 K for $x = 0.025$. The entropy corresponding to the anomaly can be estimated to be $S =$ 0.53 (0.09$R$ln2) and 0.38 J mol$^{-1}$ K$^{-1}$ (0.07$R$ln2) for $x = 0$ and $x = 0.025$, respectively, which are significantly smaller than the entropy associated with the antiferromagnetic spin ordering ($R$ln2) usually seen in the Mott transition of cuprates [25]. As shown in Fig. 3(b), the anomaly is accompanied by a small drop in $\chi$, which is observed in the charge density wave (CDW) transition due to a decrease in the electronic density of states. However, this interpretation is inconsistent with the drop in $\rho$ below the peak temperature in $C_p$ ($T_A$); the typical CDW transition leads to an increase in $\rho$, except for a few systems [26], and the structural origin of this problem remains to be resolved at present.

Figure 4 shows the phase diagram obtained by the data points of the present measurements and reference 12, together with the variation of the overall thermodynamic quantities. With increasing $x$, the anomaly at ~150 K is driven down both in temperature and entropy, and the superconducting state emerges rapidly. The superconducting state appears for $x \geq 0.035$ with a maximum $T_c$ of ~26 K at $x = 0.11$. For doping in the region of $x < 0.035$, enhancements in $T_{SF}$ and $R_W$ are observed. Together with the enhancement in the density of states estimated from $\chi$ and $C_p$, this behavior indicates an increase in the FM fluctuation in this region [27]. In the whole region of $x$, both density of states show similar $x$-dependence; there is a cusp-like maximum at $x = 0.05$. The cusp-like behavior and the larger value of $N_D^*$ than $N_D$ are similar to the critical behavior near the FM instability [27]. Therefore, the $x = 0.05$ sample may be located closest to an FM instability. These results indicate that FM spin fluctuation can induce a high superconducting transition temperature in the LaFeAsO system, in sharp contrast to high-$T_c$ cuprates driven by AF spin fluctuation.

$\chi$ and $C_p$ of LaFeAsO$_{1-x}$F$_x$ ($x = 0 - 0.025$) exhibit the characteristic temperature dependence expected for a FM spin fluctuation system. Qualitative analysis of these data leads to an experimental definition of the characteristic temperature ($T_{SF}$), the Wilson ratio ($R_W$), and the density of states ($N_D$, $N_D^*$). These values indicate that the LaFeAsO system can be considered as a nearly FM metal. Although the phase diagram obtained as a function of $x$ is analogous to the antiferromagnetism-superconductivity phase diagram found in high-$T_c$ cuprates, the observed entropy change and magnetic behavior associated with the anomaly at ~150 K are clearly different from an antiferromagnetic transition. Considering the significantly high-$T_c$ value observed in the LaFeAsO system, further efforts are certainly warranted.

We thank with Prof. K. Ishida and Dr. M. Tachibana for stimulating discussions. This work was partly supported by Grant-in-Aid JSPS Fellows (No. 19·9728 to Y. Kohama).

[1] J. G. Bednorz and K. A. Müller, Z. Phys. B **64**, 189 (1986).

[2] T. Moriya, Y. Takahashi, and K. Ueda, J. Phys. Soc. Jpn. **59**, 2905 (1990).

[3] P. Monthoux, A. V. Balatsky, and D. Pines, Phys. Rev. Lett. **67**, 3448 (1991).

[4] J. Zaanen and O. Gunnarsson, Phys. Rev. B **40**, 7391 (1989).

[5] Y. Bang, A. V. Balatsky, F. Wastin, and J. D. Thompson, Phys. Rev. B **70**, 104512 (2004).

[6] J. Schmalian, Phys. Rev. Lett. **81**, 4232 (1998).

[7] S. S. Saxena, et. al., Nature (London) **406**, 587 (2000).

[8] N. T. Huy, et. al., Phys. Rev. Lett. **99**, 067006 (2007).

[9] Y. Kamihara, T. Watanabe, M. Hirano, and H. Hosono, J. Am. Chem. Soc. **130**, 3296 (2008).

[10] (additional information) See EPAPS Document No . xxxxx for additional figures. For more information on EPAPS, see http://www.xx.

[11] H. Yanagi *et al.,* Phy. Rev. B (to be submitted).

[12] Y. Kamihara *et al.,* J. Am. Chem. Soc. **128**, 10012 (2006).

[13] T. Watanabe *et al.,* Inorg. Chem. **46**, 7719 (2007).

[14] M. T. Béal-Monod, S. K. Ma, and D. R. Fredkin, Phys. Rev. Lett. **20**, 929 (1968).

[15] S. Doniach and S. Engelsberg, Phys. Rev. Lett. **17**, 750 (1966).

[16] Y. Kohama, Y. Kamihara, H. Kawaji, T. Atake, and H. Hosono, arXiv: 0806.3139 (2008).

[17] N. E. Phillips *et al.,* Phys. Rev. Lett. **65**, 357 (1990).

[18] R. J. Trainor, M. B. Brodsky, and H. V. Culbert, Phys. Rev. Lett., **34**, 1019 (1975).

[19] P. H. Frings and J. J. M. Franse, Phys. Rev. B, **31**, 4355 (1985).

[20] M. S. Wire, G. R. Stewart, W. R. Johanson, Z. Fisk, and J. L. Smith, Phys. Rev. B **27**, 6518 (1983).

[21] W. L. McMillan, Phys. Rev. **167**, 331 (1968).

[22] S. R. Julian *et al.,* Physics B **259-261**, 928 (1999).

[23] S. Ikeda, Y. Maeno, S. Nakatsuji, M. Kosaka, and Y. Uwatoko, Phys. Rev. B **62,** R6089 (2000).

[24] C. Enss, and S. Hunklinger, Low-temperature Physics (Springer, Berlin, 2005).

[25] Y. Kimishima and H. Kittaka, Physica C **160**, 136 (1989).

[26] E. Morosan *et al*., Nature physics **2**, 544 (2006).

[27] E. Bucher, W. F. Brinkman, J. P. Maita, and H. J. Williams, Phys. Rev. Lett. **18**, 1125 (1967).

**Figure captions**

Fig. 1   (Color online) Thermal and magnetic properties for LaFeAsO$_{1-x}$F$_x$ ($x$ = 0, 0.025, 0.05, 0.11, and 0.14). (a) Temperature dependence of $C_p$. Black, blue, green, red, and purple circles correspond to the samples with $x$ = 0, 0.025, 0.05, 0.11, and 0.14, respectively. The inset shows the heat capacity difference ($C_p^*$) between superconducting samples and the non-superconducting LaFeAsO divided by temperature ($C_p^* T^{-1}$). (b) Temperature dependence of $\chi$. The data in the normal state are estimated from the slope of the magnetization curve between 1 and 2 T. The data in the superconducting state are estimated from the slope of the magnetization curve below 0.01 T. The inset of this figure shows the data below 35 K.

Fig. 2   (Color online) Detection of spin fluctuation in LaFeAsO system. (a) $C_p T^{-1}$ versus $T^2$ plot. The broken lines indicate the best fits of $C_p = \gamma_{SC} T + \beta T^3$, in which $\gamma_{SC}$ = 2.6 mJ K$^{-2}$ mol$^{-1}$, $\beta$ = 0.24 mJ K$^{-4}$ mol$^{-1}$ for $x$ = 0.05, $\gamma_{SC}$ = 2.5 mJ K$^{-2}$ mol$^{-1}$, $\beta$ = 0.27 mJ K$^{-4}$ mol$^{-1}$ for $x$ = 0.11, and $\gamma_{SC}$ = 7.7 mJ K$^{-2}$ mol$^{-1}$, $\beta$ = 0.21 mJ K$^{-4}$ mol$^{-1}$ for $x$ = 0.14. The solid curves indicate the fits of $C_p = \gamma_{SF} T + \beta T^3 + \delta T^3 (T/T_{SF})$, in which $\gamma_{SF}$ = 7.6 mJ K$^{-2}$ mol$^{-1}$ and $\delta$ = 0.23 mJ K$^{-4}$ mol$^{-1}$ for $x$ = 0 and $\gamma_{SF}$ = 9.6 mJ K$^{-2}$ mol$^{-1}$ and $\delta$ = 0.24 mJ K$^{-4}$ mol$^{-1}$ for $x$ = 0.025. The other parameters are denoted in the text. (b) $\chi$ versus $T^2$ plot for $x$ = 0 and 0.025.

Fig. 3   (Color online) Thermal and magnetic property around 150 K. (a) Excess heat capacity ($C_{ex}$) divided by temperature ($C_{ex} T^{-1}$). $C_{ex}$ is estimated by the polynomial fit of the Debye temperature [10]. (b): Temperature dependence of $\chi$ around 150 K.

Fig. 4   Upper panel: Superconductivity phase diagrams of the LaFeAsO$_{1-x}$F$_x$ system. Lower panel: Density of states and Wilson ratio as a function of F content. Lines are visual guides.

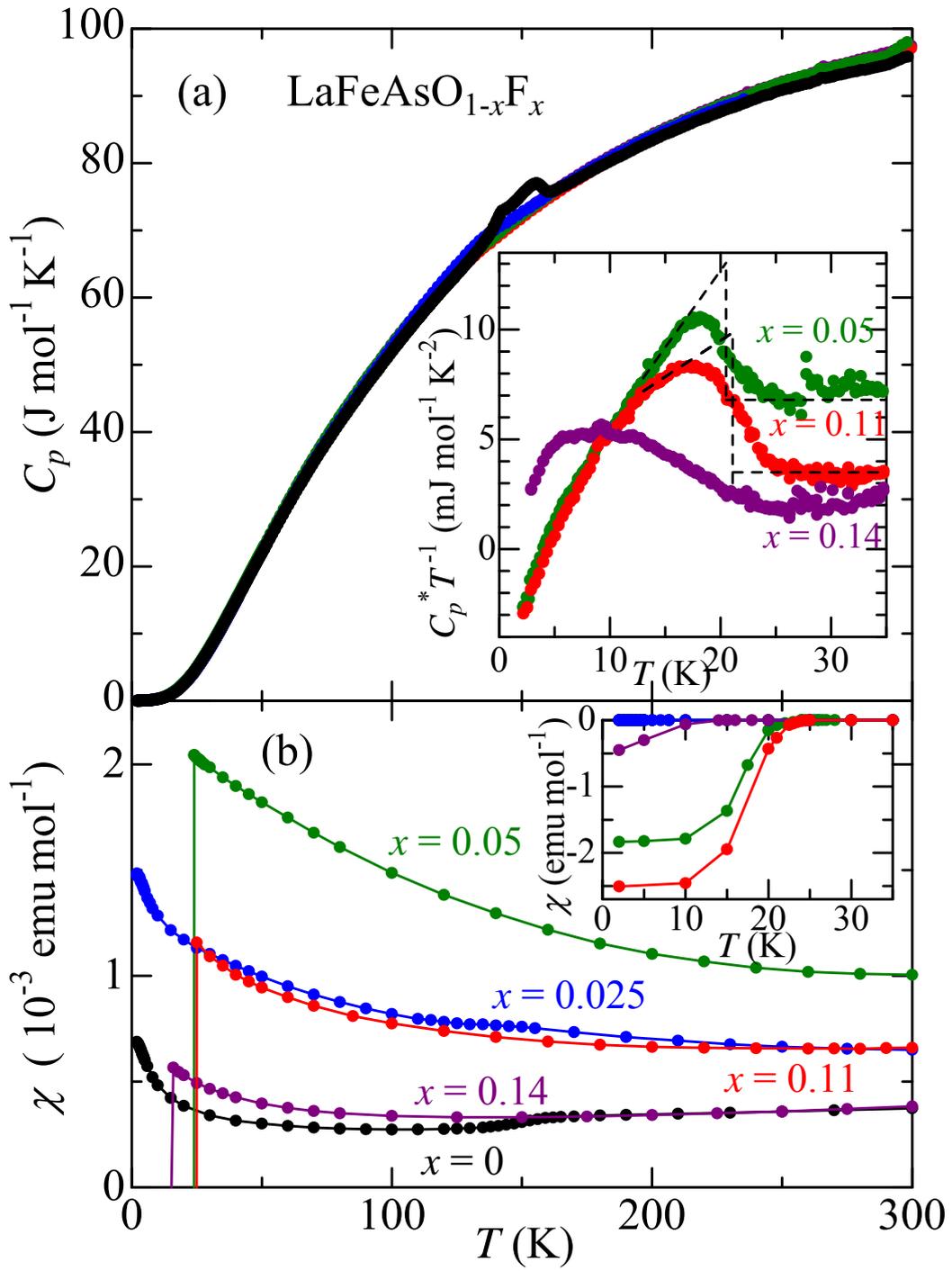

Fig. 1. Y. KOHAMA, et al.

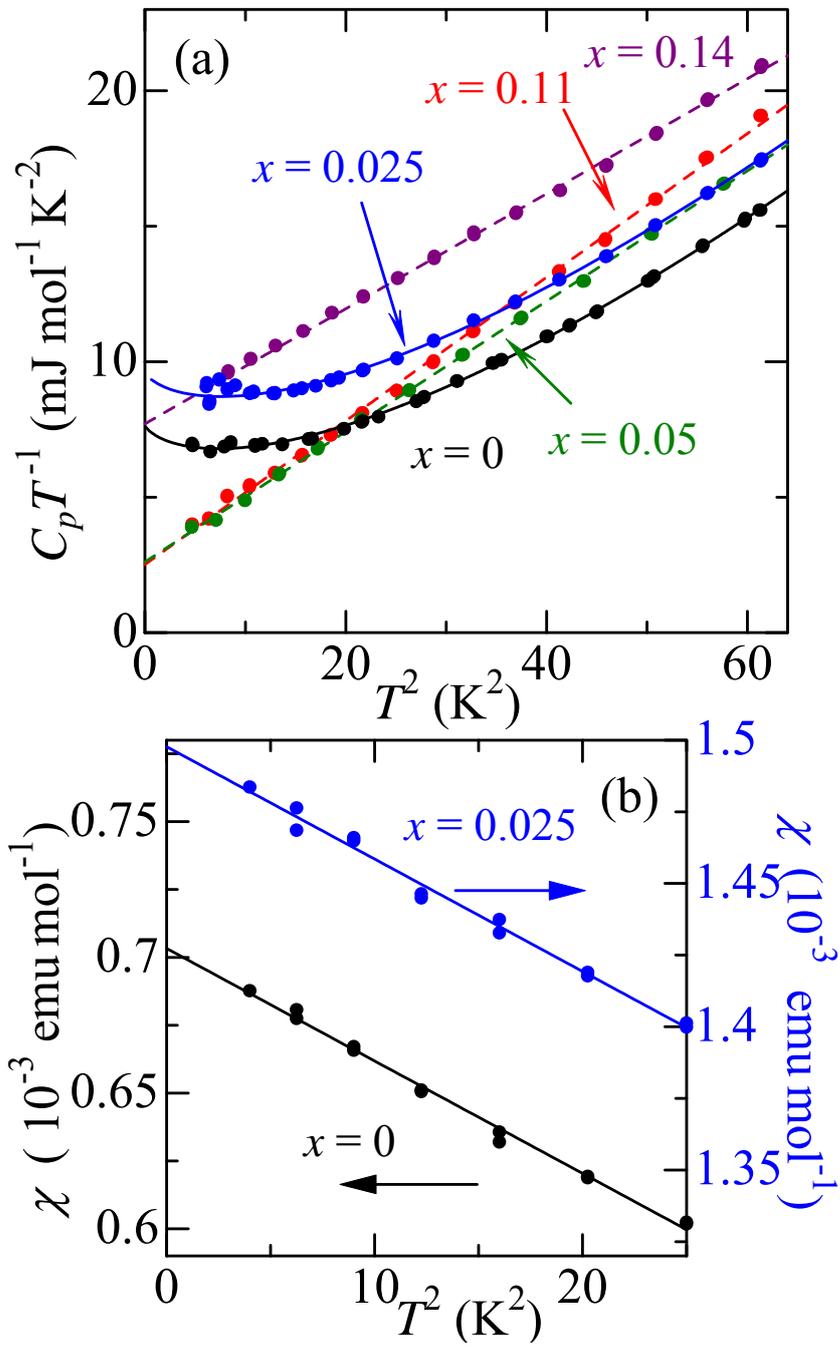

Fig. 2. Y. KOHAMA, et al.

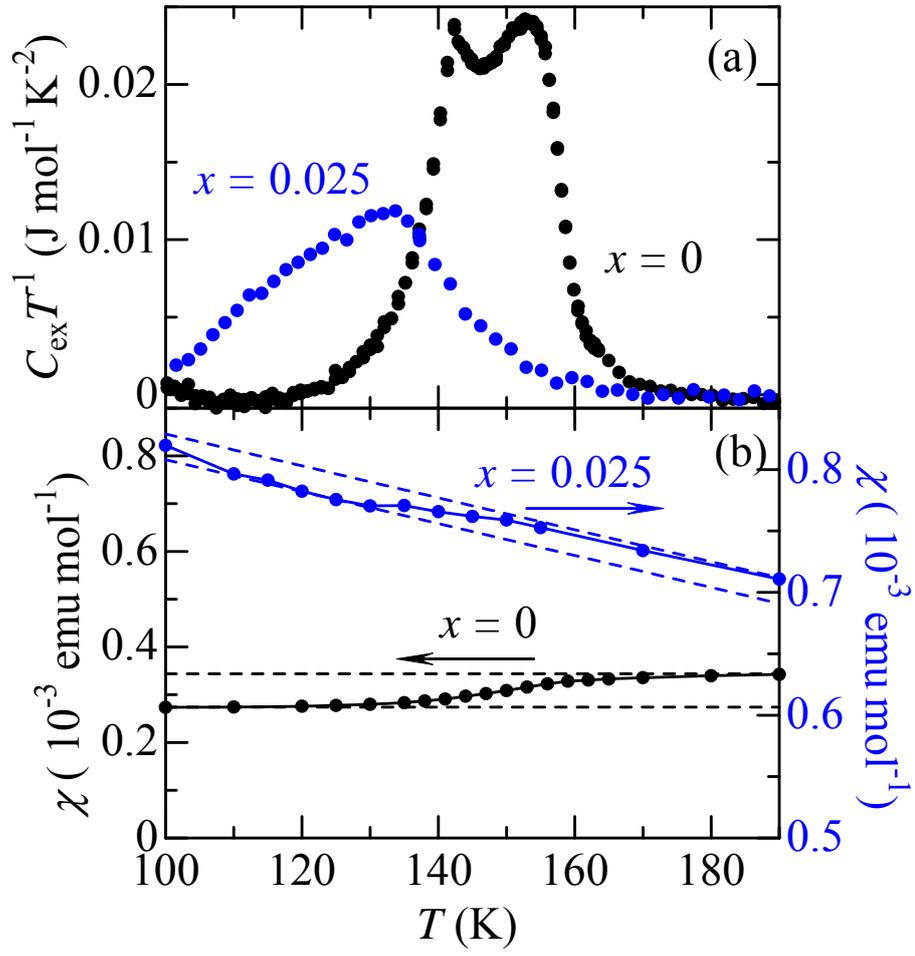

Fig. 3.   Y. KOHAMA, et al.

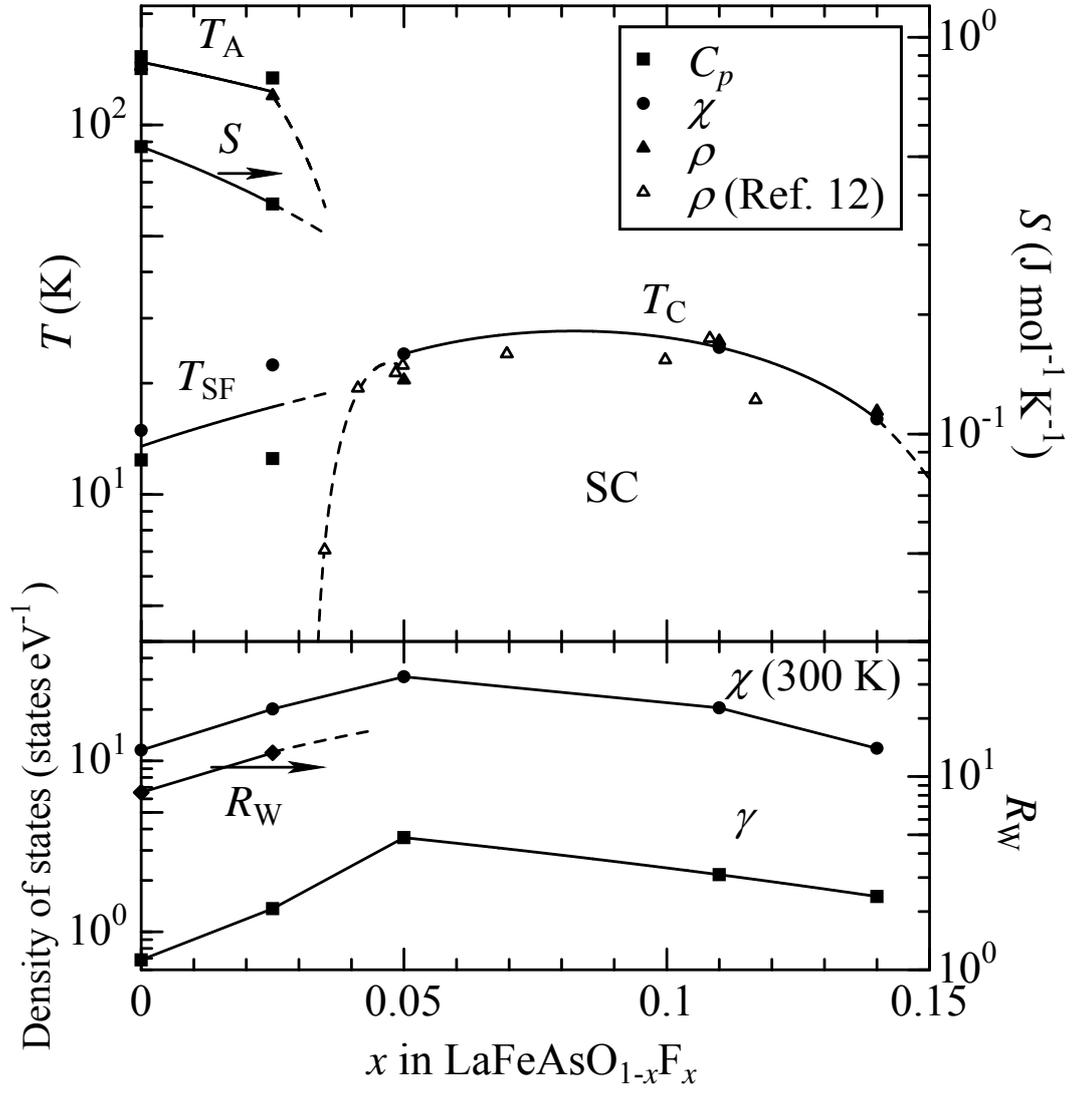

Fig. 4.　Y. KOHAMA, et al.